Subtle Sound Design

Designing for experience blend in a historic house museum

Mia F. Yates, IT-University of Copenhagen, Denmark

Anders S. Løvlie, IT-University of Copenhagen, Denmark

In this article, we present and discuss a user-study prototype, developed for Bakkehuset historic house museum in Copenhagen. We examine how the prototype - a digital sound installation - can expand visitors' experiences of the house and offer encounters with immaterial cultural heritage. Historic house museums often hold back on utilizing digital communication tools inside the houses, since a central purpose of this type of museum is to preserve an original environment. Digital communication tools however hold great potential for facilitating rich encounters with cultural heritage and in particular with the immaterial aspects of museum collections and their histories. In this article we present our design steps and choices, aiming at subtly and seamlessly adding a digital dimension to a historic house. Based on qualitative interviews, we evaluate how the sound installation at Bakkehuset is sensed, interpreted, and used by visitors as part of their museum experience. In turn, we shed light on the historic house museum as a distinct design context for designing hybrid visitor experiences and point to the potentials of digital communication tools in this context.

## 1 INTRODUCTION

There is great interest among researchers and practitioners in exploring how digital technologies may be used to facilitate new and interesting visitor experiences in museums [20,47]. Digital technologies are now used for a range of purposes, such as to increase and diversify the use of and access to collections and knowledge on-line, and thereby reaching new virtual audiences. This includes many innovative initiatives during societal lockdowns relating to the COVID-19 pandemic. Use of digital tools as part of physical exhibition design on-site has also developed rapidly over the past decade, with many museums using digital tools for creating diverse visitor engagement in exhibitions. Overall, the rapid development of digital technologies has dovetailed with an increased focus on engaging and activating the visitor in narrative and multisensory ways, leading one scholar to proclaim an "immersive turn" in the museum sector. [23]

However, implementing new technologies in historic house museums is not entirely unproblematic. Such museums often have strict preservation requirements and fragile environments that make it challenging to alter or add anything new. In addition, research points to concerns amongst museum staff that digital technologies and media might disturb visitors' perceptions of the houses as original and authentic places [11,12]. As a result, many historic house museums that do venture into using digital tools in the historic rooms limit the use to mobile based technologies such as smart phone apps or QR codes, that allow the visual and material environment to stay intact. However, the literature on technology in museums has brought up concerns about the so-called "heads-down phenomenon", referring to the fear (e.g. among museum professionals) that visitors' interaction with mobile devices may divert their attention away from the treasured objects surrounding them in the exhibition [21,28,34,45,49]. This issue reflects broader concerns in contemporary society about the dominance of smartphones and social media over our attention and face-to-face interactions [42] reflected in debates about norms for smartphone use [27,30,37,40]

For designers working with museum exhibitions this poses a challenge: How may digital technologies be used to design for innovative and immersive museum experiences, while making sure that the technologies do not dominate and steal the visitor's attention away from the physical museum exhibitions? Furthermore, how might this happen in a historic environment specifically, and in ways that don't disrupt the visitors' perceptions of the environment as 'real' or authentic?

This article reports on a research-through-design [17,51] and research-in-the-wild [35] study exploring how to design a visitor experience that brings to life some of the immaterial aspects of the history presented in the historic house museum Bakkehuset in Copenhagen - while attempting not to disrupt the experience of an original historic environment. Bakkehuset was once the home of Kamma and Knud Lyne Rahbek, two prominent members of the cultural and literary scene of the "Danish Golden Age" period in the early 1800s. They were known, among other things, as hosts of a unique social environment. In the article we report on the design of a sound installation for the house, aimed at communicating about the welcoming, informal, and cheerful social atmosphere that once filled the house, and which many of the historical guests of the house in the 1800s have described in their letters, diaries and memoirs.

Our research strategy was based on the idea of "expansive drifting" as suggested by Krogh.et.al [24] and referring to a particular way of experimenting and producing knowledge within RtD. With this research strategy, the focus is on expanding or broadening knowledge about an area of interest through multiple and diverse experiments, rather than aiming at one perfect design. In turn, experiments can be non-successive, meaning that they don't necessarily build directly on top of each other or aim towards a strict common goal, but rather explore different aspects of the topic in focus. As such, the prototype was meant as a first step in learning about the possibilities and limitations of using digital communication tools inside this historic house.

This article contributes by offering a novel approach to making immaterial heritage come to life in the museum context, through a sound installation using subtle design tactics in order to blend into the visitor experience of the historic environment. Furthermore, the study is informed by an in-depth qualitative study of the visitor experience with data collected before and after the installation of the digital experience, allowing us to assess the impact of the technology on the overall experience of the house museum.

We begin the article with situating our study within the interdisciplinary field of Museum Experience Design, by relating it to 1) digital museum experiences, as a concern of Interaction Design and Human-Computer Interaction, and then 2) Museum and Heritage Studies, concerned with the historic house museum as a distinct type of museum and design context. We continue by introducing Bakkehuset and its visitors and then move on to present our prototype development process and design choices. We then go on to present an evaluation of how visitors have experienced and used the sound installation and we discuss why and how the design expands visitors' experiences of the house and its past. In the end we reflect on the design implications for future hybrid designs in historic environments.

## 2 RELATED RESEARCH

### 2.1 Digital museum experiences

There is a wide body of research on the use of digital technologies and media in museums [20]. Research has explored the use of specific technologies or systems such as Computer Games, Virtual Reality, Augmented Reality, Tangible User Interfaces [33,34] and Artifical Intelligence [10,15]; whereas some have explored design approaches that are not tied to specific technologies [2,47]. In regards to the latter, much research within the interdisciplinary field of museum experience design rather focuses on the human-computer relationship from different perspectives, including the social and interpersonal meaning making processes [13] or affective aspects [5,8,36]. The use of screen based technology in museums has been met with some skepticism from museum professionals concerned about the before-mentioned "heads-down phenomenon": That if the museum invites visitors to use smartphone apps to enhance their visit, they might miss out on the unique collections of objects exhibited by the museum [21,28,32,45,49]. In relation to this, other scholars have explored technology-mediated "heads-up" experiences in which the technology supports and augments the visitor's experience without detracting from the physical artefacts on display [1,13,16,22,36]. Lange et al. [25] suggest a design framework for "blended experiences", aiming to facilitate digital designs that integrate well with both the museum context and the users' motivation and activities. Their model builds in part on the widely cited 'trajectories framework', which describes how artists and researchers have collaborated to create complex mixed reality experiences involving art and technology and which offer coherent experiences with a great sense of continuity [3,4]. Others use the term 'hybrid designs' to refer to a similar understanding of mixed reality designs that expand or alter a physical and sometimes original spaces through a digital layer, and with an emphasis on creating a close dialogue between the physical environment and the added digital layers [47]. In this article we also use the term hybrid, to refer to such a design.

Amongst scholars working with designing hybrid experiences in museums, the term 'seamless' design is sometimes used to refer to the careful and subtle integration of digital elements into the original historic environment. Among others, Claisse et. al. use this term in their article: "Multisensory Interactive Storytelling to Augment the Visit of a Historical House Museum"[11]. In this article, the authors zoom in on the historic house museum as a distinct context for designing hybrid experiences, and they present the development of a digital design from several principles, based on interviews with house museum experts, Claisse et.al. suggest that historic house museums have tended to hold back on incorporating digital technologies in their original house environments, and often limit the use of digital aids to pre-or post visit activities, often in separate rooms or buildings, rather than inside the historic environment. According to their expert interviews, the reasons for keeping digital elements out include museum staff's perception of digital technology as "a detached or isolated experience based on screen- or button based interfaces", which can act as a barrier for experiencing the house [11]. Understandably, such a limited perception of what digital tools might offer, can make it challenging to imagine a successful and meaningful merging of digital elements with an original house museum environment. In turn, the authors point to how digital designs for house museums should aim at a "seamless experience", in order not to interfere with the perceived authentic qualities of the house [11]. Claisse et.al present a design based on interactive storytelling where visitors take active part in an unfolding narrative, through tangible interaction with NTF tagged objects and tableaux's placed throughout the exhibition. They argue that the digital augmentation of the house provided new ways for visitors to interact with the house, such as being able to generate content and stories and to participate over a longer period of time.

In this article we build on such studies, but present a different perspective. Much of the literature on digital designs in house museums focuses on tangible interaction [11,36] and less on digital designs that visitors cannot see or touch. By creating and studying an invisible prototype based on sonic communication, this article sheds light on how digital sound can merge with the historic environment and ignite sensory and imaginary encounters with some of the immaterial qualities of the historic house.

**2.2   The historic house museum as design context**

The historic house museum is a particular kind of museum in which visitors can engage with the past in particular ways related to the sensory and imaginary immersion that the unique house environment can provide [12,31,43,50]. Some authors point to how house museums can be considered universally understood, in that they represent the concept of 'home', which is relatable in some way or form to any visitor and plays into how the visit is performed and how content is understood by visitors [12,26]. Pavoni [31] acknowledges this by discussing the recognizability of rooms and interior with their own built-in significances and narratives that influence how visitors perceive and understand the home: "The house is 'real' because it reflects a cognitive code that has been applied and tested in everyday life" [31]. Gregory and Witcomb suggest that historic house museums offer affective engagement with the past by offering visitors a chance to sense the past through its absence: "Such an affected response is heightened by the silence of the house, the absence of real life living within it. For in silence, in gaps, there is presence" [18].

In much museum visitor research emphasis has been placed on how and to whom museum experiences are meaningful, in addition to what the collections are about [48]. In line with this, emphasis has also been placed on how museums can make their collections and content accessible and relevant in both a physical and intellectual manner [6,7,38,39]. Along the same lines, house museums are also being challenged on how they present the past and help visitors understand and engage with the houses in meaningful ways that are relevant to them. Some criticize the house museums for offering a romanticized view of the past, overindulging in nostalgia and avoiding critical perspectives [18]. Author Linda Young suggests that house museums need to explore new ways of relating to visitors, rather than sticking to conventional histories. Writing about the English house museums, Young (with reference to Tinniswood) observes: "To the despair of specialists and connoisseurs, most visitors to house museums visit, not to appreciate the finer details of Tudor panelling and Georgian portraits, but to engage in creative fantasy". She continues to say that museum visitors "can be viewed as coming to mine the place for the raw materials of imaginative bricolage, and then to share the experience with their family and friends" [50]

An important aspect of the house museum visit is the perception of encountering a 'real' or authentic place. Authenticity is a common topic for house museums, since both visitors and museum staff find authenticity to be at the center of the house museum visit [11,29,44]. In relation to this, researchers have examined how house museum visitors articulate their own experiences of authenticity [29], or how curators construct authenticity [11,44]. Anna Venturini writes that:

> "Desire to see the "real thing" still motivates millions of tourists traveling to heritage sites worldwide, and several studies have proved that perceived authenticity of, specifically, the museum and its materials deeply affects visitors' impression of the overall tour, acting as a crucial driver along-side individual attitudes and expectations about the visit" [44]

In all of the ways mentioned, the historic house museum provides a unique context for designing visitor experiences, and more specifically, for designing digital encounters with cultural heritage and with the past. In this article we present a design process in close dialogue with these unique aspects of the historic house museum, including notions of home, of presence and absence and the perception of a 'real' or authentic place. We aim to explore how the careful and subtle integration of digital technologies into a historic environment can take shape and we discuss the challenges and learning outcomes that have become apparent during the process.

## 3  RESEARCH PROJECT, APPROACHES AND METHODS

This study is part of a 3-year research project (2021-2023), led by the first author, in a collaboration with The Frederiksberg Museums, comprising 4 different exhibition sites around Frederiksberg, of which Bakkehuset is one. The project is aimed at generating knowledge about the use of digital tools for museum communication about immaterial cultural heritage, through design research. The overall approach is inspired by Research through Design [17,51] in which insights emerge through the process of designing and evaluating a prototype. The project also has an element of Research-in-the-Wild [35] as the prototype was implemented and tested in the historic house museum by the visitors of the museum, offering insights on the design in the real context of an ordinary visit to the museum.

The prototype presented in this article was designed in close collaboration between researchers and museum staff through several co-creative design exercises and ongoing discussions. The study includes data gathered before and during testing of the prototype, offering a strong basis for assessing the impact of the prototype on the visitor experience. Initially, a round of interviews was carried out with 25 visitors (14 interviews). The interviews took place over a few weeks in 2021 and were carried out immediately after the museum visit had ended, in the courtyard in front of the museum. Most of the visitors interviewed had entered the museum on their own initiative, but due to low visitor numbers caused by the COVID-19 pandemic some additional local citizens were reached via posts in Facebook groups offering free access to the museum, in return for a short interview after the visit. The interviews lasted between 10 and 30 minutes and were aimed at learning about: who visited the museum, why they visited, how they experienced the house and the communication methods inside and which topics or questions they were interested in during their visits.

Additionally, another round of visitor interviews was carried out after the prototype had been finished and installed at the museum. In the second round, we spoke to 36 visitors (20 interviews) and once again the groups of interviewees consisted of people visiting on their own initiative and visitors offered free access to the museum though Facebook. The interviews lasted between 10-25 minutes each. The purpose of this second round was to learn about if and how the prototype influenced the visit, including how it might have added to or altered the visitors' experiences and meaning-making processes.

Interviewees in both rounds of interviews were evenly spread out over different age groups from 30 years and up and with variety in educational background and gender. All interviews were done just after the visitors had come out of the museum. In both rounds of interviews, the conversations were recorded and later transcribed and coded in digital software, before being analyzed.

**4   BAKKEHUSET: A SOCIAL HOTSPOT FOR WRITERS IN EARLY 1800'S COPENHAGEN**
Our study takes place in Bakkehuset (The Hill House), which was built in 1674 and is located in Frederiksberg, Copenhagen. Bakkehuset, as it was also referred to in the social circles of Kamma and Knud Lyne Rahbek, was a gathering spot for aspiring and accomplished writers, actors, and poets, who met to discuss literature, arts, politics as well as more personal topics. Kamma Rahbek became known in the literary circles of Copenhagen at this time, as an important host and sparring partner of several famous writers of the time. Her intellect and warmth drew many writers to her gatherings at Bakkehuset, including the now world famous writer H.C. Andersen. Kammas husband Knud Lyne Rahbek was himself also a writer and playwright for the theatre of Copenhagen, known also for his weekly literature publications and for helping aspiring actors and writers in the industry. All in all, Bakkehuset was a literary and cultural center in the golden age period and there are many first hand records of the warm, welcoming, cheerful, creative atmosphere of the house. As such, the museum not only represents a building and physical place, but also immaterial heritage in terms of the very particular social environment that unfolded in the house, and which later shaped parts of Danish collective memory of the golden age period and its artistic and literary activity. The museum today serves as a "literature museum" and "authentic golden age home" according to the museum website.

Apart from the physical building the museum collection consists mainly of furniture, Kammas decorative boxes and personal items, including books and letters. Most of the historic rooms are sparsely furnished and in tidily order, whereas the main living room is organized more like a tableau with a few empty tea cups and other items placed as if a small social event had taken place in the room just moments ago.

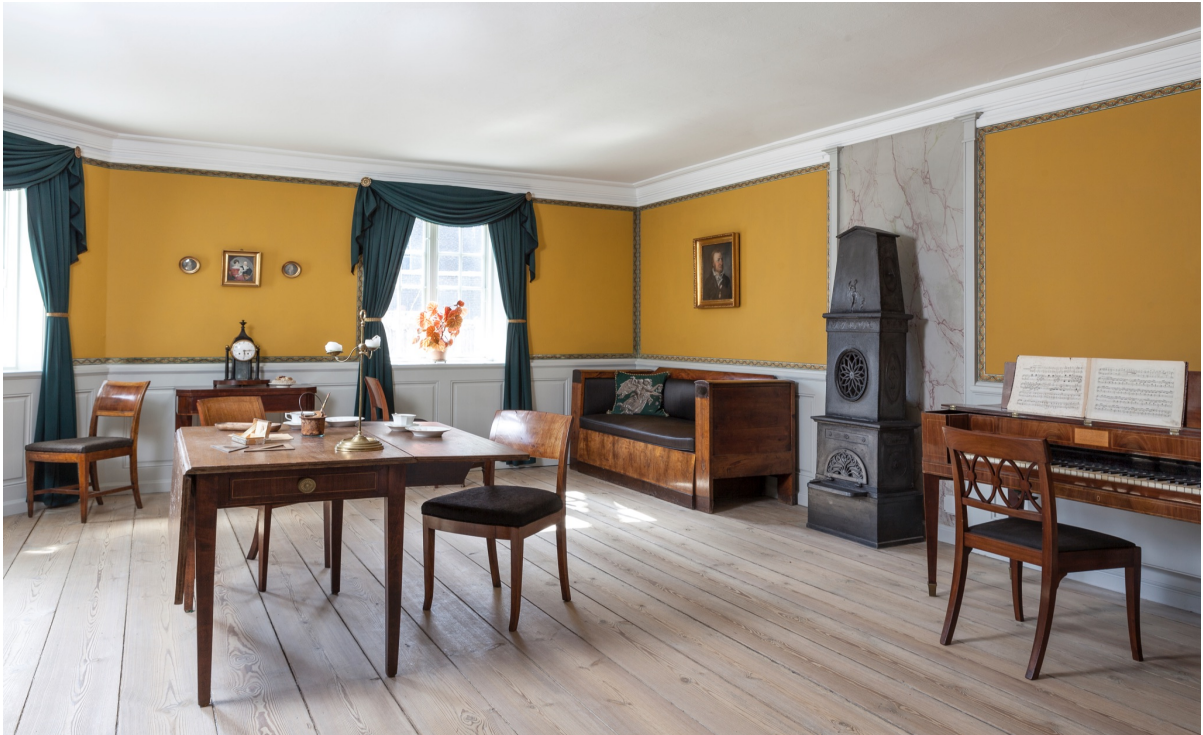
Figure 1. The living room in Bakkehuset. Image courtesy Stuart Mc.InTyre.

Bakkehuset is an example of a house museum known not so much for the building itself, as for the people who lived in it and for the social and creative practices that unfolded in the house, much in line with the type of museum referred to by Linda Young as a "hero museum". Young describes this type of house museum as one where someone important lived or passed through and of which she writes:

> "The management of intangible meanings - reasons for which a place is significant, even sacred-
> is a central function in the hero's house. This is a site infused with the aura of the famous person
> and, even if its meanings are unhistorical, where such values continue to inspire the community,
> they warrant the acknowledgement of preservation.' [50]

Bakkehuset doesn't only represent one hero, but a whole group of some of Denmark's most well-known writers and poets of the golden age period. This social group and the cultural milieu and atmosphere surrounding them is the reason why the house is still famous, hundreds of years later. In turn Bakkehuset, just like many other historic homes, needs meaningful ways to communicate about the social environment and atmosphere, which cannot necessarily be pointed to inside the house. Although there is an openness towards digital potentials, the museum has so far held back on adding any explicit communication (digital or other) inside the house, due in part to the strict preservation requirements and in part to the fear of disturbing visitors' (and staff's) perceptions of the house as original and authentic. Communication efforts inside the historic rooms have therefore been limited to a few text labels and text-based folders provided at the front desk. During the time of our study however, the museum was in the process of developing a mobile-based audio guide with additional information about the rooms and people of the house. The audioguide was not fully developed or available at the time of our study, and is therefore not included.

## 5  WHY DO PEOPLE VISIT BAKKEHUSET?

Knowing how visitor experiences are highly influenced by pre-existing knowledge, expectations, social and cultural backgrounds as well as the social context of the visit [14] we started our design process by doing a series of interviews with current visitors to the house museum, about their motivations, knowledge, interests, and experiences during their visits at Bakkehuset. Here we briefly present some insights from the interviews that were important for the forthcoming design process.

First, we found that most of our interviewees had very similar motivations for visiting Bakkehuset. The most common reason for visiting was to be able to imagine and immerse oneself in the past. Visitors expected to encounter an old and original home and to be able to see and imagine how everyday life was lived in the early 1800s. Visitors often described the house as a "time capsule" and most assumed that the house was more or less the same as when the couple lived in the house and also found it important whether objects in the house were in fact 'real' in the sense of being preserved original artefacts. Second, many visitors mentioned an appreciation of the "calm atmosphere" in the house, described as a sense of visual and sonic calmness, which importantly, was not disrupted by any modern elements or digital devices. Visitors appreciated that there were no text labels or screens bombarding them with information and that, apart from a few other guests and distant sounds from the city, the house was quiet, which also allowed them to have intimate conversations with their group. Third, although many visitors appreciated the calmness of the house, many also mentioned how the house seemed almost too empty and orderly to be a 'real' authentic home. They were missing the material traces of the life that had unfolded in the house and requested a lot more furniture, objects and messy rooms, resembling scenes from what they considered a 'real' home. Some expressed that the housed lacked "personality" or that it felt "lifeless" because of its emptiness. As such, although the concept of home can be universally understood, as mentioned above, in this case, the environment inside the house didn't necessarily correlate with visitor's personal perceptions of what a 'real' home is supposed to look like. Fourth, we also learned that very few visitors were interested in, or had any knowledge about, the literature of the time or the literary works of residents and guests to the house. This was interesting to us, since the museum considers itself in part a "literary museum". Almost no visitors asked for information on literature or the literary scene of the time, but did however show an interest in the few famous writers they knew had visited the house, such as H.C. Andersen. However, the request was rather for knowledge about his social activities in the house. What happened, where did he sit, what did he do in the house? Questions related to the everyday activities in the house and his role in those, rather than any literary topics, relating also to Linda Youngs observation, that house museum visitors are not necessarily interested in factual knowledge about the objects or furniture, but in actively imagining how they were once used by people. [50] These findings initially allowed us, as designers, to better understand how the house was being used and understood by visitors and in turn what to have in mind when developing a design for the house.

## 6  DEVELOPING THE IDEA: PROCESS AND DESIGN CHOICES

In order to establish a solid grounding in the views of the museum stakeholders, we set up a design team consisting of three museum staff along with two design researchers. Together we organized a first workshop aimed at 1) identifying the social milieu and atmosphere we wanted to communicate about to visitors, and 2) discussing how authenticity should play into the design. Both parts of the workshop included classic design activities such as oral and written brainstorms and affinity mapping exercises. The second part of the workshop, focusing on the term authenticity, was meant to open up a joint discussion about the distinct nature of the house and its meanings. During a structured activity, we discussed when and how something can be perceived to be authentic, based on personal experiences shared by each participant in the workshop. The activity sensitized us to a clearer understanding of the term, and a more nuanced consideration of the different ways in which an experience can be felt, sensed, or understood, as being authentic. Our discussions lead us to clearly distinguishing between something being *factually original*, in the sense of belonging to the right historical time period or to the house or its residents, and then something being *experienced as authentic*, which is rather a subjective and individual matter. In museum practice, the word authenticity is often used in ways that blur the boundary between these two understandings, which is why other researchers have also previously discussed similar distinctions in relation to authenticity in museums and tourism [29,46]. On top of providing us with a clearer understanding of and joint terminology for the word authenticity, the affinity mapping exercise led us to identify several common and (to us) important characteristics of authentic experiences (based on our own subjective examples). We used these common characteristics as a basis for formulating several design guidelines, that we would keep in mind during the forthcoming design process. This included considerations of 1) how the design would allow for different sensations (auditory, visual, olfactory etc.) to come together meaningfully and create a sensory immersion for visitors; 2) how visitors knowledge about and expectations of the house and its residents would play into perceptions of something being authentic or not; 3) how we could create a design which included or related to the material environment and even indexical traces (in a Peircean manner) of the house residents - such as for instance original handwritten letters or personal items, and 4) how our design could incorporate something unfiltered or flawed, which was also a recurring topic when sharing our personal examples of authentic experiences.

Next, we did an ideation workshop meant to provide a co-creative space for the design team members to brainstorm further and make decisions, leading to a more concrete idea. We used several analogue props for physically developing the idea. Props included a printed out A3 paper map of the historic house and a large number of words printed on paper and cut out to be handled and placed on the map. During our discussions, sound was a recurring theme. We talked about how laughter, discussions, singing, reading aloud, small talk and making jokes were all part of the oral culture that somewhat constituted the social atmosphere of the house. From the written memoirs and diaries of people who were part of the social scene at the time, we had detailed descriptions of what exactly was once said, song and read in the house. In the end, we decided to turn these written testimonies into a sonic echo of the past, by means of creating a sound installation, mimicking the social activities which had once unfolded in the house, and more specifically, in the living room where most of the social activities had taken place around the dining table (see figure 1). Using sound as the primary communication method also corresponded well with the aim of designing for a 'seamless' and invisible experience and would allow us to examine intangible aspects and potentials of digital communication tools.

### 7   PRODUCING AND INSTALLING THE SOUND, SUBTLY AND SEAMLESSLY

After jointly deciding on the rough idea for a sound installation during the workshop, we went on to carefully write a manuscript over the next few weeks. The manuscript consisted of five different social scenarios, played out one after the other and lasting 15 minutes in total. The scenes conveyed three named characters (one woman and two men), sitting at the dining table or by the piano. The scenes consisted of poetry readings, short discussions about contemporary poetry or politics, small talk about everyday life, or collective singing or playing the piano. Any music played or sung was either written by Knud Rahbek himself, or it was other music of the time that we knew they could have possibly known and liked. The manuscript was acted out by professional actors, and supplied by carefully chosen sound effects. Piano melodies were played and recorded by a professional piano player, with a piano sound corresponding with the piano standing in the living room of the house. Everything was recorded in a studio by professional sound technicians, and in the end we had a 15-minute sound file which we wanted to play in a loop in the living room.

In developing the manuscript and choosing the sounds, our goal was to create a realistic interpretation of what it might have sounded like inside the house, according to the written testimonies, and with emphasis on the social interactions that had unfolded there. Bearing in mind the ideas of authenticity from our earlier workshops, we deliberately designed for flaws in the sound. For instance, we included mistakes in the singing or piano playing - some deliberate, and some accidental - with the intention of making it sound like a group of friends getting together casually in a living room, rather than a perfect music recording. Human sounds like coughing and other 'natural' bodily sounds were also included. In such ways we attempted to create a sonic space that would fit common experiences of what takes place in a living room between friends.

We also chose sounds in direct relation to the specific interior and materials of the room. For instance, we made sure to record sounds of someone sitting down on a *wooden* chair or of glasses being moved around on a *wooden* table, relating to the materials of the dining table and chairs in the museum. We placed copies of original handwritten letters and original books of poetry on the dining table, allowing the sounds of someone reading or turning a page to correspond with such objects. We also placed fresh flowers and fruit on the table, in order to include olfactory sensations in the experience. Overall, we tried to make sure that different sensory aspects of the room would fit together carefully. We intended the sound to not only merge with the visual experience but also to enhance it, and vice versa. We made sure that no wires or other traces of the technology were visible, using only battery-powered speakers and digital units which we hid inside the furniture. Any sounds relating to the piano would come from a speaker hidden inside the piano, whereas sounds of talking or singing would come from a hidden speaker underneath the dining table. Each speaker was connected to iPads through Bluetooth and the audio files were played from a freeware audio player. As such, we did our best to create a 'seamless' design where the different sensory dimensions would fit together as organically as possible.

We discussed extensively the exact volume of the sound. Initially when installing the sound, members of our design team disagreed slightly about how high the volume should be. Some felt that the volume should be somewhat subdued, to avoid that the sound would be too dominating and disturbing to visitors in other rooms. Others believed that the volume needed to be as close to natural as possible for it to seem realistic. However, once we tested out the installation at a lower volume, the entire team agreed that this felt too unnatural. For instance, a piano being played enthusiastically or someone collectively singing drinking songs together are loud activities. Turning the volume down on such scenes made the sound feel fake. Thus the volume was adjusted to be as close to the natural level as possible.

## 8   INTERNAL TEST AND REDESIGN

As an initial assessment of the design we conducted an internal evaluation, gathering our design team in the living room and listening to the full 15-minute sound file play out in the room. We quickly agreed that the parts of the sound which had dialogue for 2-3 minutes in one go were too demanding to listen to. Although the sound quality was excellent, listening to the dialogue played out in the room was difficult. In fact, any dialogue of more than 15-20 seconds at a time didn't work well when played out into the room, as opposed to listening to it via headphones, which is what we ourselves had done up until that point. The sound was interrupted by other sounds in the room – of steps on the squeaky wooden floors or someone speaking, which made it hard to focus on what was being said. Also, the type of dialogue we had chosen, like poetry readings and political discussions, were hard to follow without having any frame of reference and not recognizing the topics or names being discussed. It became apparent to us that the ability to relate to the content and meanings of such narrative sound changes tremendously, depending on whether you listen through headphones that have an isolation factor by filtering out other sounds and making it easier to concentrate, or whether you listen in a room where the entire sonic landscape of the house and other people in it become part of the listening experience. In addition, we also considered what visitors in our first round of interviews had said about the importance of the calm atmosphere of the house, and the opportunity to talk to one another during their visits. Therefore, we decided to cut the sound file into five small separate scenarios of between 1-3 minutes each, which then faded into complete silence for 3-4 minutes in between each scenario. This turned the audio-file into a total of about 30 minutes, including the silent breaks. The parts of the dialogue we decided to cut out were the longer dialogues about literature or politics, since we knew that this was less important to visitors. Instead, we kept the shorter dialogues of a maximum of 15-20 seconds at a time, such as small talk about everyday life and practicalities around the dining table, and then mixed with piano playing, singing and other sound effects. This design choice changed the prototype from communicating about a *literature* milieu, to communicating about a more *general* social environment in a living room space, and as such the meaning of the sound installation also changed. The design choice was an important step in allowing the design to better integrate seamlessly with the visitor flows and social aspects of the museum visit.

**9  A SENSE OF HUMAN PRESENCE AND ATMOSPHERE: VISITOR EXPERIENCES**

While we hoped that most visitors would experience the sound installation as a positive augmentation of the house, we also expected that some visitors might find the installation misplaced or disruptive in the historic environment. After installing the sound installation, we carried out a qualitative evaluation over a few weeks, in which a researcher was stationed outside the museum exit and approached visitors for an interview as they exited the museum. The visitors were told that the interview was about their overall experience of the museum visit and with a particular focus on communication methods. Deliberately, visitors were not told that we were specifically interested in their experiences of the sound installation, as we wanted to understand how the installation influenced their overall experience, and thus were interested in seeing if and how they would bring that aspect up without being prompted for it. Also, we did not place any written information about the sound installation inside the museum, and neither did the front desk staff say anything about it to visitors. We wanted to know if the sound installation could stand alone without any additional information, or whether this would be confusing to visitors, who might have questions about what they were hearing.

Most visitors did mention the sound installation on their own initiative, and the vast majority found the installation to add positively to their experience of the house, in different ways. They described the overall experience of the sound using a diverse range of words like "atmospheric", "beautiful", "lovely", "fun", "cozy" and "nice". Defying our expectations, only 2 of the 36 interviewees expressed a dislike towards the sound installation, which we will come back to.

In the visitors' descriptions of hearing the sound in the living room, the most dominant response was that the sound created a sense of a human presence and through that - a sense of atmosphere.
The sounds gave visitors the impression that there was a human presence in the room, and that someone was there, or had been there. Some expressed it as follows:

> "I thought it was cool how there was music and a little talking and a few sounds, so that you sense some life in the house. I liked that"

> "I noticed that there was singing. So, you got the sense that someone had sat there and entertained themselves with singing. And then there was some chatting about something, almost like it was festive, at one point"

The sounds we had chosen clearly mimicked recognizable human actions like drinking, eating, singing and evidently these sounds triggered images in the visitors' minds, of homely and familiar scenarios, also reflecting Pavoni's [31] observations about visitors' pre-existing cognitive codes related to notions of 'home'. Visitors actively imagined what once took place, by combining the sounds with their own mental images and memories and thus created their own personal narratives of how life in Bakkehuset had once unfolded. This finding is in line with previous research into multisensory museum experiences, pointing to how imagination and memory is fundamentally multisensory and as such why different sensations can trigger and ignite rich experiences for visitors, exactly because it supports their ability to use their imagination (Ward et. al 2019). Relating to this, another visitor said:

> "You can really imagine how they were living in here, and also; we just heard a little song in there, in the main living room. So, you can actually *see* the people there. (...) You can *see* them sitting around the table, making music and so (…) Well, it helps me to imagine how life was back then"

As mentioned, many interviewees had found that the house was almost too empty and orderly to be a real home. However, some suggested that the sounds counterbalanced this emptiness, by adding to or filling the room, implying a sense of a material quality to the sound. One visitor said:

> "I'm glad it is there because it adds some life to these rooms that I'm surprised to find otherwise kind of lifeless".

The sense of a human presence that visitors described, was also sometimes connected to a momentary understanding of the sounds as real. Many visitors were initially surprised by the sounds and thought they were coming from a real person in the room. Two friends describe such a situation:

> Visitor 1: "I suddenly asked my friend: what did you do? Like, where did that come from?
> Visitor 2: "Yes, because it came from inside the room, and then I thought to myself, 'now you feel his presence'. And I looked around like, 'is someone in the room'? (…) like, 'is someone standing in the corner'. That's what it was like. But it was very peaceful."

> Visitor 1: "A lovely experience"
> Visitor 2: "Yes it was"

Such perceptions of the sounds as being real-time sounds, was not an intended effect. Although we had aimed at making it sound as realistic as possible, we had also expected that visitors would immediately recognize the sounds as added and performed sounds coming from a speaker. However, it seemed that especially the non-narrative and isolated sounds, such as someone coughing or humming, or a chair being pulled, would sometimes be experienced as real. Also, several mentioned how, when standing in a different part of the house and hearing the piano being played or someone singing, they would get drawn to the living room, and would get curious about who was singing and what was going on. Going back to our discussions of the volume level, this confirmed that having the sound at natural volume, worked well. Although it was loud at times, it didn't disturb visitors, but rather drew them to the living room because they assumed they were hearing real-time sounds. Interestingly, although visitors soon after realized the staged nature of the sounds, they still attributed a human or lively quality to the sound. Overall, this unintended 'real-time effect' seemed to add positively to the experience as visitors explained how it made them alert and surprised them in a positive way.

Additionally, visitors described how the sense of a human presence they felt in the house added or enhanced a more overall sense of an atmosphere. Here are some examples of such statements:

> "I like that it's not a story, because you don't need to follow what they say. I don't like when it's a whole story, because then you must stand there and wait until it starts again, and you must hear it all. Here, it's just an atmosphere, some talking. And I thought that was nice."

> "I heard some speaking and then some piano playing and singing. And I think that's quite good, because it creates an atmosphere".

> Visitor 1: "I heard some… I don't really know what they talked about, it was more of a mumbling in the background and someone pouring something, and there was singing and someone talking about how he had to reach Copenhagen and about the weather or something.
> Visitor 2: "(…) it worked well I thought. I mean, it enhanced the atmosphere."

Historically speaking, learning outcomes and gaining new knowledge has been a major focus point in museum communication and exhibition planning and equally so in evaluations of visitor experiences [19]. In this case however, none of the visitors explained that they had gained new information or *learned* something specific, from hearing the sound. Rather, we had set the *stage* for visitors to sense something, but in the end, what they actually sensed and comprehended depended widely on how they combined their experiences in the house with their personal knowledge and imagination. Philosopher Gernot Boehme, who wrote on creating and perceiving atmospheres said that:

> "(…) atmosphere itself is not a thing; it is rather a floating in-between, something between things and the perceiving subjects. The making of atmospheres is therefore confined to setting the conditions in which the atmosphere appears". [9]

In line with Boehme's description, the social atmosphere that we had set out to communicate about, took shape in the meeting between the physical space (including sound), and the imaginative interpretation processes in the minds of the individual visitors. What they sensed, was something intangible yet very sensory present.

Returning now to the two people who clearly stated that they did not enjoy the sound installation, one of them, a man in his 70's, could not explain why he didn't like it. The other, a man in his 30´s with a background in literature and history, explained that he did not visit the museum to step into the past or to imagine anything. He came for facts about this period in history, about the writers of the time and to learn more about their literary works. In this way, the man was an exception, in terms of motivation, expectations and prior knowledge, and this relates directly to why he found the sound installation useless. He understood the entire house as a replica of sorts, something fundamentally inauthentic. In line with this, he also found the sound installation "unconvincing" and did not mention anything about sensing an atmosphere, human presence or anything else positive related to the sound. This visitor was a great example of how prior knowledge, expectations and interests always play into how visitors perceive and experience museums and their communication designs differently. At Bakkehuset however, most visitors do enter with the expectation of encountering a 'real' and old home. They want to imagine and comprehend the past and they aim at actively using whatever communication tools are present, to make this happen.

## 10  DISCUSSION: DESIGN CHOICES AND IMPLICATIONS

The relative success of this sound design for a house museum seems to be related to some key design choices, which we will now discuss in relation to three related topics: visitor motivations, open-ended sound, and social flows.

To begin with, we want to reflect on how visitor motivations played into the design. Most visitors to Bakkehuset enter with an expectation of encountering an old and 'real' home and wanting to imagine how life was lived in the past. Knowing this and continuously making design choices that would cater to this expectation was the most fundamental and important design choice leading to a communication design which according to visitors "suited" the house. Because what this really meant was that the design suited their *expectations* and *purposes* of visiting. The design allowed them to more easily imagine how life was lived in the house, which in turn echoes Linda Young (and Tinniswood) in her argument that most visitors to house museums visit to engage in creative fantasy [41,50]. In our design, we made it possible for visitors to engage in that creative fantasy.

Secondly, the open-endedness of sounds was important. The choice and duration of sounds allowed visitors to get just a sense of something going on, rather than being presented with a complete storyline. Subtle sound effects like squeaky floors, subtle humming, coughing, or pouring of wine would give visitors little impressions or hints, without being too detailed or descriptive. They prickled at visitors' imagination, but did not provide any direct messages or answers, and as such were not designed for a specific didactical outcome. The diffuse, non-factual and non-literal nature of the sounds allowed them to be interpreted by visitors to fit their personal understandings of 'home', of the house and of the past, exactly because there was no fixed storyline. The sounds of piano playing and collective singing communicated something slightly more specific to visitors, since there were more words, and since a few guests also recognized the lyrics or festive melodies. However, since the music sounded like a casual impulsive sing-a-long amongst friends, and with no factual information, visitors seemed to just enjoy the sentiment of the music. They could float in and out of it as they pleased, rather than stopping and listening to all of it, or questioning what they were listening to. Surprisingly not a single visitor mentioned having any unanswered questions about the what, why, who and how of the sound installation. This indicated that additional information wasn't needed in this case. What visitors got from the sound was *hints* of something, and a sense of something, rather than a literal description. As such, we see how an open-ended and atmospheric sound design like this can counter the 'heads down' phenomenon and provide an alternative to mobile-based solutions. In this case, the visitor experience was not an isolated or detached one, as some museum staff fear when considering digital solutions, according to Claisse et. al. [11]. Rather the digital dimension provided a sensory and imaginative enrichment of the visit. Again, this is especially relevant for historic houses and sites as distinct contexts for cultural heritage encounters, where one wants visitors to be attentive to, and present in, the visual and material environment surrounding them.

Thirdly, an important aspect of why visitors had a positive experience of the sound was due to how the installation integrated with visitors' social flows through the house. Many visitors explained that because the sound was only set up in one single room of the house, and because it had pauses of complete silence, it was neither overwhelming, dominating or intrusive. The frequent pauses in the sound allowed visitors to still sense the calmness and quiet of the house, just as they were still able to talk to each other in between sounds, or even over the sounds – due to the non-literal and subtle nature of the sounds. Had we chosen to set up our first version of the prototype with a non-stop 15-minute audiofile, we presumably would have gotten a very different response to the installation. This first non-stop version would have demanded that visitors stayed in one place for a long time and would have most likely made it difficult for visitors to share the experience, just like different groups of visitors would quite possibly have disturbed each other in their listening. In this second, more subtle version of the prototype, visitors felt like they were able to use the house and journey through it as they pleased, and as such to remain in control of their journey through the house. These more practical aspects of integrating the design into the social use and flows of the house were vital in terms of making the design integrate 'seamlessly' with the overall museum visit.

To sum up, what turned out to have a particularly strong impact on our design being experienced by visitors as "suiting" was: the way the design supported visitors' motivations and expectations; the open-ended nature of the sounds that let visitors 'fill in' the blanks with their imagination; and the way the design allowed visitors to remain in control of their visits and share the experience with their group.

**CONCLUSION**

In our study we set out to explore how to subtly and seamlessly integrate digital technologies into a historic house museum. This meant using digital technologies in such a way that they would not dominate the experience or disrupt the visitors' perception of the house as 'real' or authentic. The aim was to also understand whether a seamless digital design might add positively to the experience and comprehension of the historic house and its past.

The overall result is that it is possible to utilize digital technologies for communication inside a historic house museum, in ways that are experienced as positive and meaningful to visitors. The concern of designers and museum staff, that adding a digital layer might possibly disturb visitors' perceptions of the house as 'real' or authentic, was not confirmed. The overall result was that visitors did not find the digital installation either disturbing, unfitting or otherwise wrong for the old house. On the contrary, the study has shown us that visitors can handle the merging of the old and original with the new and digital. Most visitors found the digital installation suiting for the house and considered it a positive expansion, by allowing them to sense a human presence and an atmosphere, and through this, to better imagine life unfolding in the house. These findings point to how hybrid museum experiences, such as a digital augmentation of a historic place, can offer valuable encounters with some of the immaterial and intangible qualities of such a place. Based on this study, historic house museums do not need to limit themselves to pre-and post-visit use of digital technologies, or strictly mobile-based technologies, which has been suggested is often the case. Rather the study points to valuable potentials of digital designs that subtly trigger sensations and imagination, without being too literal or directly interactive.

Our design was traditional in its aim to cater to visitor's expectations and to the idea of the past as something different and distant, that we can look back on and 'visit'. But there are many other ways of designing for the distinct context of the historic house museum environment which might rather challenge, question or alter perceptions of home, time, historic space or lived life, in ways that might still be considered 'seamless'. Such perspectives, within the 'seamless' design approach to experience designs in historic house museums, would be interesting and relevant to study further.